\documentclass[twocolumn]{book}
\usepackage[dvips]{graphicx,color}
\usepackage{makeidx,universe}

\makeauthorindex

\BookTitle{Proceedings of the XXIX PHYSICS IN COLLISION}

\CopyRight{\copyright 2009 by Universal Academy Press, Inc.}

\begin{document} 

\pagenumbering{arabic}

\chapter{%
{\LARGE \sf
Observation of Ultra-high Energy Cosmic Rays
} \\
{\normalsize \bf 
Radom\'{\i}r \v{S}m\'{\i}da$^1$ } for the Pierre Auger Collaboration\\
{\small \it \vspace{-.5\baselineskip}
(1) Institute of Physics, Academy of Sciences of the Czech Republic, 
      Na Slovance 2, Prague 182~21, Czech Republic
}
}




  \baselineskip=10pt 
  \parindent=10pt    

\section*{Abstract} 

The measurement of ultra-high energy cosmic rays is an unique way to study particle interactions at energies which are well above
the capability of current accelerators. Significant progress in this field has occurred during last years, particularly due to the measurements 
made at the Pierre Auger Observatory. The important results which were achieved during last years are described here. Also future plans for the study
of cosmic rays are presented.

\section{Introduction} 

Cosmic rays are observed in a wide range of energy, but their sources remain unknown. The well known prominent source of cosmic rays, 
but only those with the energy below a few $10^9\ \mathrm{eV}$, is the Sun. We can expect that also other stars emit cosmic ray particles 
into interplanetary space. There are several indirect measurements which indicate that expanding shells of supernova remnants accelerate 
particles up to very high energies. The maximum energy which can be achieved 
by particles in these remnants is about $10^{15}\ \mathrm{eV}$ for protons and up to $10^{18}\ \mathrm{eV}$ for iron nuclei.
The origin of ultra-high energy cosmic rays (UHECRs) -- we will call this name for cosmic rays with the energy above $10^{18}\ \mathrm{eV}$ 
-- is still unknown, but results of the measurements indicate that sources are located outside our Galaxy. 

The search for UHECRs sources must take into account the magnetic fields which are measured in interstellar and also intergalactic space.
Cosmic rays as charged particles are significantly deflected during their propagation because of their electric charge.
Therefore their arrival directions do not point back to their sources. Our knowledge of magnetic fields is not sufficient for the 
calculation of trajectories of measured particles. The main attention is thus focused on the most energetic cosmic rays
which may propagate through magnetic fields along almost linear trajectories. 

Moreover, there is a strict limit on a distance of any source of cosmic rays with the energy above $\sim 4\times10^{19}\ \mathrm{eV}$ 
as was shown by \cite{Greisen,Zatsepin}. Such energetic cosmic rays can not survive their propagation over the distance longer than $\sim100\ \mathrm{Mpc}$. 
This lead to a restriction on the number of possible sources which should be present as a suppression in the flux of cosmic rays. 
This predicted suppression in cosmic ray energy spectrum is called Greisen-Zatsepin-Kuzmin (GZK) cut-off.

The methods of UHECRs measurement are described as the first in this article. Then a brief description of current experiments follows
with special attention to the Pierre Auger Observatory and the results of its measurement. Finally, future plans in cosmic ray measurements 
are mentioned.

\section{Cosmic Ray Measurement} 

Cosmic rays with the energy higher than $10^{15}\ \mathrm{eV}$ cannot be measured by satellite or high-altitude balloon experiments
because of their low flux. Therefore, they are measured indirectly by the observation of extensive air showers from the ground. 
The extensive air shower is a bundle of secondary particles initiated by a primary particle which arrived into the atmosphere. 
The energy of the primary particle is distributed among secondary particles in subsequent collisions in the atmosphere.

There are two detection techniques used for the observation of extensive air showers. The sampling of secondary particles arriving 
at the ground is the first one. The surface detector consists of a net of scintillators, water Cherenkov tanks or muon detectors. 
An arrival direction and energy of a primary cosmic ray can be reconstructed from timing information of triggered stations 
and lateral distribution of measured signal, respectively. The main advantages of the surface detector are an automatic operation, 
almost 100\% duty cycle and simple calculation of its exposure. On the other hand, the energy reconstruction is based on models of hadronic
interactions extrapolating data from laboratory experiments at lower energies. The uncertainty in hadronic interactions constitute 
a large contribution to a systematic error of the energy reconstruction. 

The second technique is based on a measurement of the isotropic fluorescence light emitted by air nitrogen molecules which were excited by secondary
particles of the extensive air shower. The longitudinal development of the extensive air shower is measured by a set of fluorescence telescopes. 
Unfortunately, the fluorescence detector can be operated only on clear and moonless nights limiting its duty cycle to about 12\%
of a year. The measured signal, after correction for an attenuation due to Rayleigh and aerosol scattering and for Cherenkov light, is proportional 
to the number of fluorescence photons emitted in the field of view of each triggered pixel. A Gaisser-Hillas function \cite{Gaisser}
is used to reconstruct the shower profile which provides a measurement of the energy of the extensive air shower deposited in the atmosphere. 
To derive the primary energy, a correction to missing energy carried into the ground by muons and neutrinos must be made. This correction
is however small and its uncertainty is less than 4\% of the energy of primary particle. The energy reconstruction is more precise than 
in a case of the surface detector, 
because a fluorescence yield for various atmospheric conditions can be measured in a laboratory. However, the uncertainty in the fluorescence yield 
is still an important source of systematic uncertainty and therefore a large effort is directed into a precise measurement of the fluorescence 
yield by laboratory experiments. Other important sources of the systematic uncertainty come from the absolute calibration of telescopes, 
absorption and scattering in the atmosphere and reconstruction methods.

\begin{figure}[t]
 \begin{tabular}{c}
  \begin{minipage}{1.0\hsize}
   \begin{center}
     \includegraphics[width=.98\textwidth]{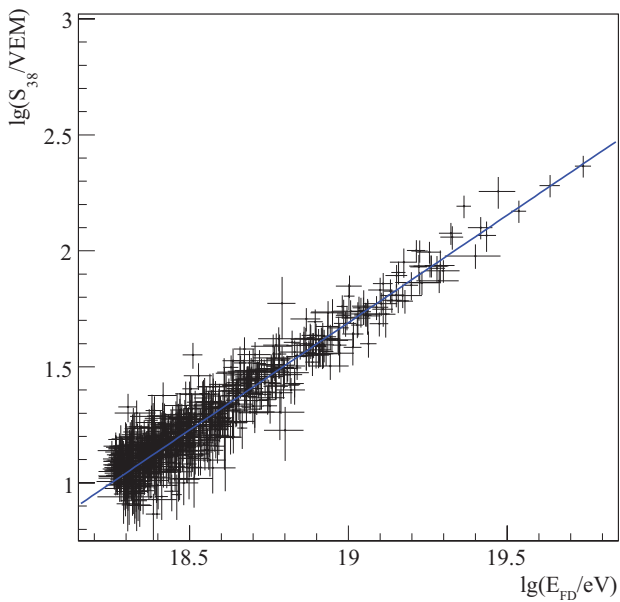}
      \caption{
      Correlation between an energy estimator of the surface detector (S$_{38}$) and the energy reconstructed from fluorescence detector data (E$_{FD}$)
      for extensive air showers measured by the Pierre Auger Observatory \cite{Auger2008}. The full line is the best fit to the data.}
    \label{fig:SD-figure1}
   \end{center}
  \end{minipage}
 \end{tabular}
\end{figure}

A significant improvement of the reconstruction can be achieved by using the data measured by both detection techniques for same extensive air showers.
A hybrid detector can be well calibrated, has well known aperture, excellent angular resolution (about 0.2$^{\circ}$) and low energy threshold.
The first hybrid detector is the Pierre Auger Observatory and it has measured a clear connection between an energy estimator measured 
by the surface detector and the energy measured by the fluorescence detector as is shown in Fig.~\ref{fig:SD-figure1}. In such a way the energy 
can be attributed to all UHECRs measured only by the surface detector.

\section{Current Experiments} 

\begin{figure}[t]
 \begin{tabular}{c}
  \begin{minipage}{1.0\hsize}
   \begin{center}
     \includegraphics[width=.98\textwidth]{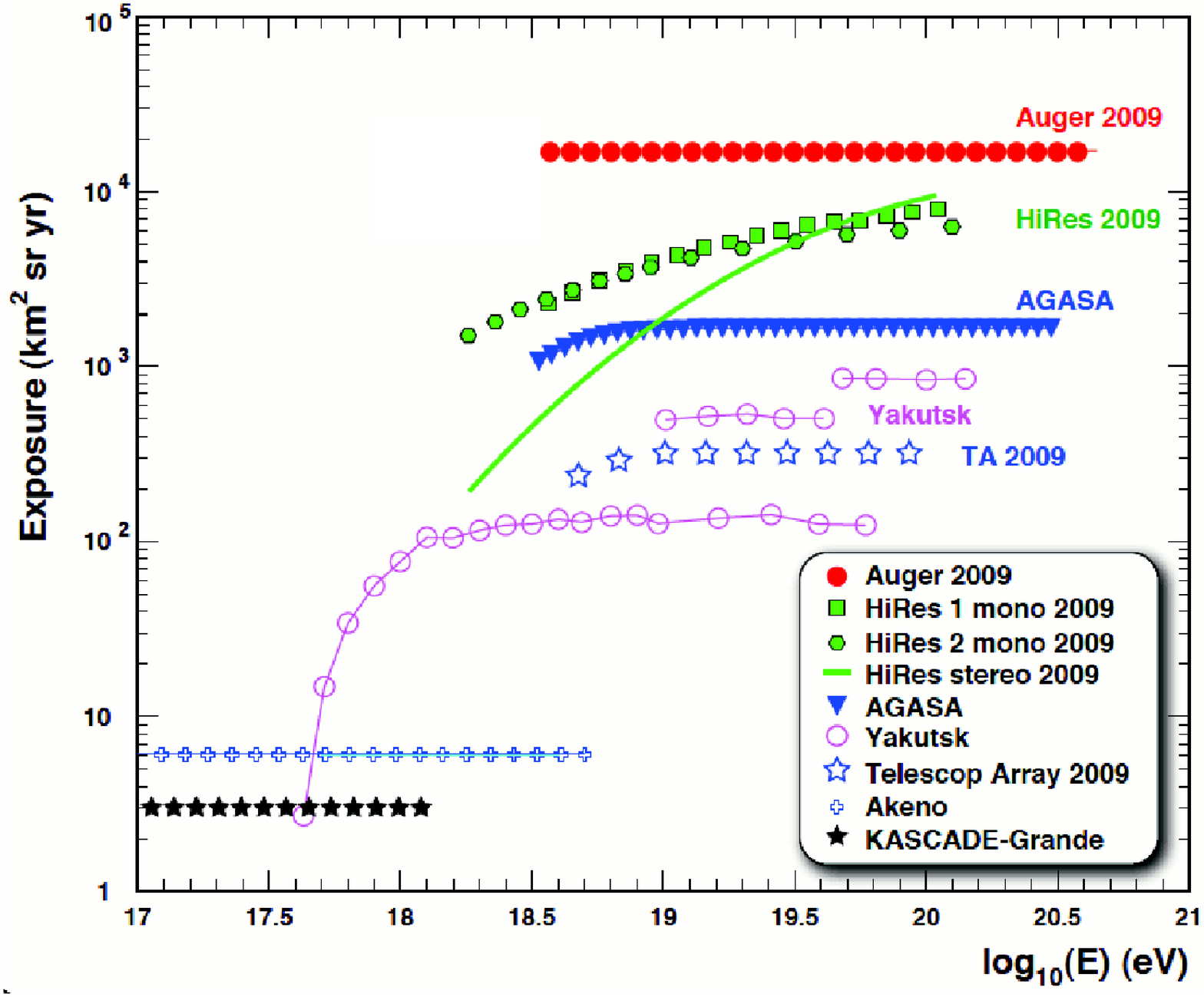}
      \caption{
      The exposures of several cosmic ray experiments as a function of measured energy \cite{Kamp}.}
    \label{fig:Exposures}
   \end{center}
  \end{minipage}
 \end{tabular}
\end{figure}

There are currently three working observatories of UHECRs in the world. Two of them are located in the northern hemisphere -- Telescope Array 
in Utah, USA \cite{Nonaka} and Yakutsk in Russia \cite{Knurenko}. The Pierre Pierre Auger Observatory is situated in western Argentina \cite{Auger2004}, 
but its second part is planned in the northern hemisphere in Colorado, USA \cite{Harton}. The exposures of current and also of some former 
observatories of UHECRs are shown in Fig.~\ref{fig:Exposures}.

The Pierre Auger Observatory and also Telescope Array will be described here. Both of them are hybrid observatories, but they differ
in the type of the surface detector. The Pierre Auger Observatory has already published its results, which are discussed
in following sections. Up to now, there was no result presented by the Telescope Array collaboration.

\subsection{Pierre Auger Observatory} 

The Pierre Auger Observatory is located in the southern hemisphere in the western Argentina province of Mendoza 
close to the city of Malarg\"{u}e (69$^{\circ}$~W, 35$^{\circ}$~S, $1\ 400$~m a.s.l.) \cite{Auger2004}. The construction of this 
largest cosmic ray observatory in the world was finished in June 2008. 

\begin{figure}[t]
 \begin{tabular}{c}
  \begin{minipage}{1.0\hsize}
   \begin{center}
     \includegraphics[width=.98\textwidth]{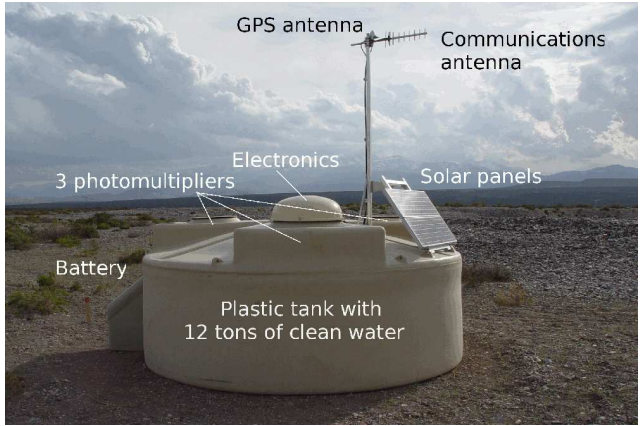}
      \caption{Picture of a water tank at the Pierre Auger Observatory. The names of the various components of the system are shown.
}
    \label{fig:SD}
   \end{center}
  \end{minipage}
 \end{tabular}
\end{figure}

{\it The surface detector} is an array of more than $1\ 600$ water Cherenkov detectors spaced at a distance of 1.5~km 
and covering a total area of $3\ 000$~km$^2$. Each detector is a plastic tank of cylindrical shape with an area of 
10~m$^2$ and height 1.2~m filled with purified water \cite{Suomijarvi}. 
Each tank is operating completely autonomously and its components are shown in Fig.~\ref{fig:SD}.

The surface detector measures the front of the shower as it reaches the ground. The tanks activated by a cosmic ray shower record 
the particle density and the time of arrival. The flux of cosmic ray muons provides a continuous monitoring of the surface detector. 
The aperture achieved with the surface array for zenith angles less than 60$^{\circ}$ is $7\ 350$~km$^2$~sr. The detection 
efficiency at the trigger level reaches 100\% for cosmic ray events with the energy above $3\times10^{18}$~eV.
By including events with larger zenith angles (up to 80$^{\circ}$) in the analysis, the aperture is increased by 30\%. 

The events recorded by the surface detector are reconstructed using the arrival time and the signal size from the shower particles reaching 
the tanks. The magnitude of the signal at 1~km from the intersection of the shower axis with the ground is $S(1000)$, measured in 
units of vertical equivalent muon (VEM) -- average signal size given by a muon crossing the tank vertically. The total signal is estimated
from the lateral distribution function fit as a size parameter of the shower. Two cosmic rays of the same energy, but incident 
at different zenith angles, will yield different values of $S(1000)$ due to the attenuation of the shower in the atmosphere. 
The attenuation curve can be obtained from measured data.

With the subset of events detected in hybrid mode (simultaneous measurement with both surface and fluorescence detectors) the link 
between $S(1000)$ for given zenith angle and the primary energy can be established independently on simulations using data from 
the fluorescence detector which provides calorimetric energy measurement (see Fig.~\ref{fig:SD-figure1}).

The arrival direction of the primary particle is inferred from the relative arrival times of the shower front at different surface 
detector tanks. The angular resolution improves with measured energy and zenith angle because of greater number of triggered stations. The surface 
array has the angular precision better than 1$^{\circ}$ in case of measurement of extensive air shower by at least four tanks.

\begin{figure}[t]
 \begin{tabular}{c}
  \begin{minipage}{1.0\hsize}
   \begin{center}
     \includegraphics[width=.98\textwidth]{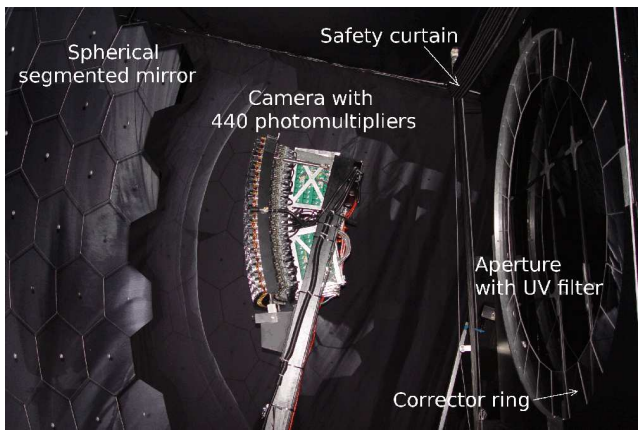}
      \caption{
      Picture of a fluorescence telescope of the Pierre Auger Observatory. The main components of optical system are indicated. Light goes 
      through UV filter with a corrector ring in an aperture and is reflected by a spherical mirror onto camera with photomultipliers. A safety 
      curtain in the aperture is used in case of emergency.
     }
    \label{fig:FD}
   \end{center}
  \end{minipage}
 \end{tabular}
\end{figure}

{\it The fluorescence detector} comprises four observation sites located atop small elevations on the perimeter of the surface detector array
\cite{AugerFD}. 
Six independent telescopes, each with the field of view of 30$^{\circ}$ times 
30$^{\circ}$ in azimuth and elevation, are located in each fluorescence detector site. The telescopes face towards the interior of the array 
so that the combination of the six telescopes provides 180$^{\circ}$ coverage in azimuth. The telescopes are housed in climate-controlled 
buildings. Nitrogen fluorescence light enters through a large UV filter window and Schmidt optics corrector ring. The light 
is focused by a spherical mirror onto a camera of 440 pixels with photomultiplier light sensors (see Fig.~\ref{fig:FD}). Light pulses 
in the pixels are digitized every 100 nanoseconds, and a hierarchy of trigger levels culminates in the detection and recording of cosmic 
ray air showers.

The calibration \cite{Bauleo,Caruso} of the fluorescence telescopes is done using accurately calibrated light sources and a cylindrical diffuser
that illuminates the camera uniformly. It is an end-to-end procedure that takes into account the transmission of the filter, the reflectivity 
of the mirror and the response of the camera photomultipliers.

A complex equipment for monitoring the atmosphere has been installed on the site of the Pierre Auger Observatory. 
This system, based on the LIDAR technique and on steerable laser beams, provides continuous information on the attenuation of the fluorescence 
light due to Rayleigh and aerosol scattering along the path from the shower to the telescopes \cite{BenZvi,Valore}. Moreover, the presence of clouds
is monitored by cloud cameras and actual temperature, humidity and pressure are measured by radiosondes \cite{Keilhauer}. These measured data
are routinely used in event reconstruction.

\subsection{Telescope Array} 

The Telescope Array project is located in the high desert in Utah, USA and its construction was finished in March 2008 
\cite{Nonaka}. It consists of a ground array of 576 scintillation counters deployed with 
a 1.2~km spacing, and three sites with the fluorescence detectors overlooking it. 

Unlike at the Pierre Auger Observatory the scintillation detectors were used at the Telescope Array. 
Each unit of the surface detector consists of two layers of plastic scintillators of 3~m$^2$ and 1.2~cm thick. 
There are 2 photomultipliers each connected with fibers to a corresponding layer.
The ground array is 100\% efficient above the energy of $10^{19}\ \mathrm{ eV}$ and has the aperture of about 
$1\ 500$~km$^2$~sr. 

One of the fluorescence detector stations located at the north of the array is a transferred station from the experiment HiRes. 
Other fluorescence detector stations at west and east side of the array were newly developed. They consist of 12 telescopes 
and the field of view of each covers $28^{\circ}$ vertically and $108^{\circ}$ horizontally.
Similarly as at the Pierre Auger Observatory the response of individual components of the detector are monitored and calibrated.
The fluorescence detector is running since November 2007. Facilities for atmosphere monitoring are also ready to be operated.

\section{Energy Spectrum} 

The shape of the energy spectrum of cosmic rays with the highest energies can be formed by several mechanisms. Particularly very important 
is the end of the cosmic ray spectrum, where a rapid decrease of cosmic ray flux is predicted by the GZK mechanism. A comparable 
decrease can be also caused by losses of acceleration efficiency in (still unknown) sources. A superposition of cosmic ray spectra from 
different sites of origin (e.g. galactic and extragalactic) can play crucial role in spectrum formation.

Different methods have been used by the Pierre Auger collaboration to measure the energy spectrum of primary cosmic rays. The data presented 
here refer to showers with zenith angle below 60$^{\circ}$. The analysis of more inclined showers requires a more complex and sophisticated 
treatment (see \cite{Vazguez}). The energy spectrum was obtained from surface detector measurements and also from hybrid data, where at least one
triggered water tank is required for an event measured simultaneously also by the fluorescence detector. The fluorescence detector has intrinsically
the capability to measure showers at energies lower than the surface detector and this enables the feature of the spectrum called ankle to be studied.

\begin{figure}[t]
 \begin{tabular}{c}
  \begin{minipage}{1.0\hsize}
   \begin{center}
     \includegraphics[width=.98\textwidth]{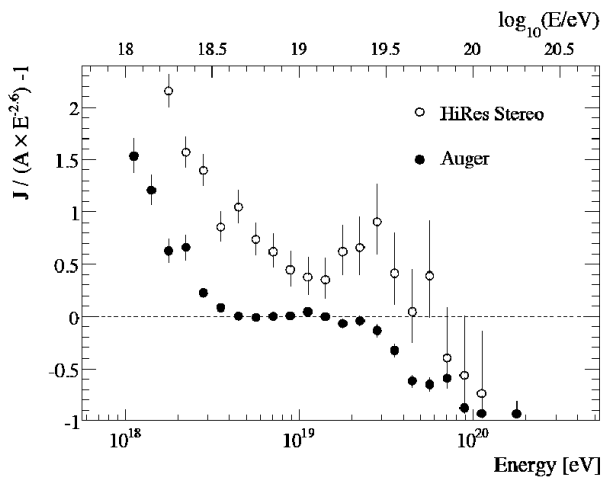}
      \caption{The fractional difference between the combined energy spectrum of the Auger Observatory and a spectrum with 
      an index of 2.6 \cite{Schussler}. Data from HiRes experiment are shown for comparison \cite{Hanl}. The two features -- the ankle and 
      the GZK suppression -- are visible.}
    \label{fig:Spectrum}
   \end{center}
  \end{minipage}
 \end{tabular}
\end{figure}

The combined spectrum measured by the Pierre Auger Observatory is shown in Fig.~\ref{fig:Spectrum}. It is based on data measured since January 2004 
till end of 2008. The total exposure of the surface detector was $12\ 790$~km$^2$~sr~yr during this period with an uncertainty of 3\% \cite{Schussler}. 
The energy calibration, as based on the fluorescence calorimetric measurement, is the same for both methods and therefore the two spectra can be 
combined together. However, the energy of the hybrid data has a statistical uncertainty of about 9\% while the energy of the surface detector data 
has 17\% uncertainty and therefore the surface detector data have to be unfolded before combination. The surface detector unfolded spectrum and 
the hybrid spectrum were found to be consistent within their uncertainties in the overlapping region.

Cosmic ray fluxes measured by the Pierre Auger Observatory and one of the previous experiment HiRes are compared in Fig.~\ref{fig:Spectrum}.
The difference between them is most likely to be attributed to the fact that the two experiments use a different energy calibration. 
In fact, applying to the data a constant relative shift of the energy scale of 25\% would essentially bring the two sets of data in agreement.

There are two changes of spectral index in the energy spectrum: the ankle at the energy of $4\times10^{18}$~eV and the suppression 
above energies one magnitude higher. The suppression is statistically very significant being at the level of more than eight standard deviations. 
Although it is consistent with the prediction of the GZK mechanism, effects caused by cosmic ray sources themselves cannot be yet excluded.

Fluorescence detector uncertainties dominate systematic uncertainties of the energy determination in the results from the Pierre Auger Observatory.
Total systematic uncertainty is 22\% at present \cite{Giulio}.

\section{Upper Limit on Flux of Photons} 

Neither photons nor neutrinos were detected above the energy of $\sim10^{14}$~eV yet. Both of them are stable and neutral particles which are not affected 
by magnetic fields during their propagation in space and therefore they point back to their sources. The successful detection of ultra high 
energy photons or neutrinos could help to solve the whole mystery of cosmic rays and lead to the discovery of point sources or indicate a need 
for a new physics. The detection of small flux of ultra-high energy photons could among others confirm the GZK mechanism.

Showers initiated by UHE photons develop differently from showers induced by nuclear primaries. Particularly, observables related to the development 
stage of measured air showers (such as the depth of shower maximum $X_{max}$) provide good sensitivity to identify 
primary photons. Photon showers are expected to develop deeper in the atmosphere (larger $X_{max}$) than hadronic ones.
Photon showers also contain fewer secondary muons, since photoproduction and direct muon pair production are expected to play 
only a sub-dominant role. The air shower maxima can be directly observed by the fluorescence detector. Moreover, two simple but robust observables 
can be extracted from data measured by the surface detector. These observables provide good discrimination
between photon and nuclear primaries. They are the radius of curvature of the shower front and the risetime at given core distance.

\begin{figure}[]
 \begin{tabular}{c}
  \begin{minipage}{1.0\hsize}
   \begin{center}
     \includegraphics[width=.98\textwidth]{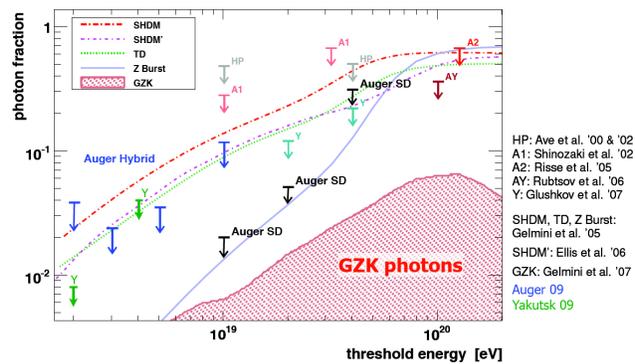}
      \caption{The upper limits on the fraction of photons in the integral cosmic ray flux measured by previous 
      and current experiments \cite{Kamp}.
      Also predictions from top-down models and predictions of the GZK photon fraction are shown. The current limit given by hybrid and surface 
      detector data measured by the Pierre Auger Observatory already excluded majority of top-down models.
      }
    \label{fig:Photon}
   \end{center}
  \end{minipage}
 \end{tabular}
\end{figure}

The Pierre Auger Observatory has the sensitivity for a detection of photons \cite{Auger2009}. Measured upper limit on the fraction of photons 
in the integral cosmic-ray flux is shown in Fig.~\ref{fig:Photon}. The limit ruled out many top-down models of cosmic-ray origin 
(i.e. cosmic rays coming from decays of superheavy particles) and in the future it could reach the level 
of GZK photons (i.e photons produced during interaction of UHECRs with the cosmic microwave background radiation). 
The corresponding photon fluxes are sensitive to source features (type of primary, injection spectrum, distance to sources etc.) and to propagation 
parameters (extragalactic radio backgrounds and magnetic fields).

\section{Upper Limit on Flux of Neutrinos} 

There are good reasons also for the searching for ultra-high energy neutrinos. Their observation should open a new window to the universe since 
they can give information on regions that are otherwise hidden from observation by large amounts of matter. Moreover, neutrinos 
are not deviated by magnetic fields and hence they maintain the direction of their production places. The existence of ultra-high energy 
cosmic rays of energies exceeding $10^{19}$~eV makes it most reasonable to expect neutrino fluxes reaching similar energies. 
Although the origin of cosmic rays and their production mechanisms are still unknown, neutrinos are expected to be produced together with 
the cosmic rays and also in their interactions with either interstellar and intergalactic matter or the background radiation fields during propagation. 
Unfortunately there are still many unknowns concerning cosmic ray origin 
and neutrino fluxes remain quite uncertain.

The neutrino detection method focuses on a search for very inclined young showers. The neutrino events would have 
a significant electromagnetic component leading to a broad time structure of detected signals in contrast to nucleonic-induced showers. 
The surface detector of the Pierre Auger Observatory has a satisfactory discrimination power against the larger background of nucleonic
showers over a broad angular range \cite{Tiffenberg}.
The Earth-skimming tau neutrinos are also expected to be observed by detecting showers induced by the decay of emerging $\tau$ leptons, after 
the propagation of neutrinos through the Earth \cite{Auger2009n}.

Up to now, no neutrino candidate has been observed in the data. The upper limit is still almost one order of magnitude above the GZK
neutrino predictions. The Pierre Auger Observatory will keep taking data for about 20 years over which the bound will improve by more 
than an order of magnitude if no neutrino candidate is found.

\section{Chemical Composition} 

The knowledge of the mass composition of ultra high energy cosmic rays is extremely important for the study of cosmic rays.
For example the hardening of the cosmic ray energy spectrum at the ankle can be explained either by the transition of galactic heavy 
nuclei to extragalactic protons or by a distortion of the extragalactic proton spectrum due to energy losses (see e.g. \cite{Berez}). 
Moreover, a measurement of the cosmic ray composition is essential to understand the nature of the flux suppression observed 
above $4\times10^{19}\ \mathrm{eV}$. 

Due to the low flux of cosmic rays at these energies the composition can not be measured directly 
but has to be inferred from observations of the extensive air showers.
The atmospheric depth $X_{max}$ at which the longitudinal development of the extensive air shower reaches its maximum is a measure of how 
fast the energy of the primary particle is transferred to electromagnetic sub-showers. The average of the shower maximum $<X_{max}>$ and 
particularly its change per decade of energy are sensitive to the particle composition. Another important feature are shower-to-shower fluctuations, 
which are expected to decrease with the number of nucleons in primary nucleus and to increase with the nuclear interaction length.

The fluorescence detector of the Pierre Auger Observatory can be used to measure with good resolution the shower longitudinal profile and the depth 
at which the shower reaches its maximum size \cite{Bellido}. Proton showers penetrate deeper into the atmosphere and have wider $X_{max}$ 
distributions than heavy nuclei. The atmospheric depth $X_{max}$ where an air shower reaches its maximum size is measured shower-by-shower 
with a resolution of 20~g/cm$^2$ at the Pierre Auger Observatory. The change from protons to heavier composition is observed also in the measured
fluctuations of $X_{max}$.

The comparison with results from the previous experiment HiRes \cite{Belz} shows agreement within their uncertainties at the highest energies. However, the results published 
by HiRes experiment focused only on the $<X_{max}>$ as a function of energy, but did not include a study of the energy evolution 
of the $X_{max}$-fluctuations and could not reach far above $10^{19}\ \mathrm{eV}$ because of limited statistics. The mass composition 
interpretation of the measured quantities depends on the assumed hadronic model. There is a problem that uncertainties on the predictions 
of hadronic models are unknown, because each model is based on an extrapolation of the physics from lower energies.

\begin{figure}[]
 \begin{tabular}{c}
  \begin{minipage}{1.0\hsize}
   \begin{center}
     \includegraphics[width=.98\textwidth]{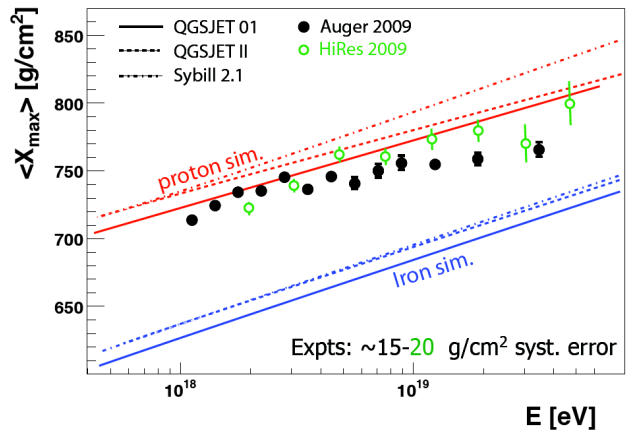}
      \caption{
The average $X_{max}$ as a function of the energy for HiRes stereo \cite{Belz} and Pierre Auger Observatory hybrid 
data \cite{Bellido} compared to proton and iron predictions using different hadronic interaction models (solid 
curves) and different models of UHECR origin \cite{Kamp}. There are still no results above the GZK energy.}
    \label{fig:Composition}
   \end{center}
  \end{minipage}
 \end{tabular}
\end{figure}

\section{Anisotropy of Arrival Directions} 

Searches for anisotropies have been done in sky regions, where some candidates for UHECRs sources can be located. An attractive candidate 
is the Galactic Centre region, since it houses some objects that are candidates for powerful accelerators. Neutrons could be produced
due to interactions of accelerated particles with ambient matter. The distance of the Galactic center from the Earth is 8.5~kpc 
and a significant fraction of neutrons with energies higher than $10^{18}$~eV will survive their path to the Earth. As neutral particles 
the neutrons will point back to their site of origin, which in this case is around the position of the Galactic Centre. 
The exposure of the position of the Galactic Centre at the Pierre Auger Observatory is higher than that of previous experiments. 
However, the results of the Pierre Auger Observatory do not show significant excesses for a point-like source or overdensities for larger 
angular windows around the Galactic Centre \cite{Auger2007}.

\begin{figure}[]
 \begin{tabular}{c}
  \begin{minipage}{1.0\hsize}
   \begin{center}
     \includegraphics[width=.98\textwidth]{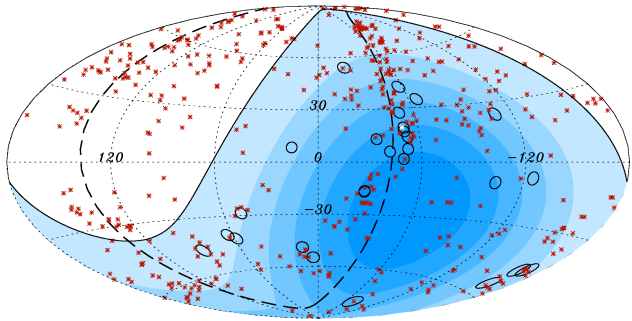}
      \caption{
      The celestial sphere in galactic coordinates with circles of radius 3.1$^{\circ}$ centered at the arrival directions of the 27 cosmic rays 
      with highest energy detected by the Pierre Auger Observatory \cite{Auger2007s}. The positions of the 472 AGN with redshift less than 0.018 (i.e. distance less than 
75~Mpc) are indicated by red asterisks. The solid line represents the border of the field of view (zenith angles smaller than 60$^{\circ}$). Darker color 
indicates larger relative exposure. Each colored band has equal integrated exposure. The dashed line is the supergalactic plane. Centaurus A, one of our
closest AGN, is marked in white.
}
    \label{fig:AGN}
   \end{center}
  \end{minipage}
 \end{tabular}
\end{figure}

One of the most important results found in the study of arrival directions was presented by the Pierre Auger Collaboration in 2007 
\cite{Auger2007s,Auger2008s}.
Using data collected between 1 January, 2004 and 31 August, 2007, the Pierre Auger Observatory has reported the evidence of anisotropy 
in the arrival directions of cosmic rays with energies exceeding $6\times10^{19}$~eV. The arrival directions were correlated with 
the positions of nearby objects from the 12th edition of the catalog of quasars and active galactic nuclei by Veron-Cetty and Veron \cite{Veron}.
In original analysis 20 from 27 measured events were found closer than $3.1^{\circ}$ from positions of AGN lying within 75 megaparsecs (see Fig.~\ref{fig:AGN}).
This degree of correlation led to the rejection of the hypothesis of an isotropic distribution of these cosmic rays with at least 
a 99\% confidence level from a prescribed a priori test. The observed correlation is compatible with the hypothesis that the highest-energy 
particles originate from nearby extragalactic sources whose flux has not been substantially reduced by interactions with the cosmic background 
radiation. AGN or objects having a similar spatial distribution could be possible sources.

Since then 31 additional events were measured and eight of them have arrival directions within 
the prescribed area of the sky (i.e. around the positions of selected AGN). This update of the analysis of the correlation with AGN, with nearly twice 
the previous exposure was done, neither strengthens nor does contradict the earlier result \cite{Hague}. The anisotropy of UHECRs arrival 
directions at the highest energies is still observed.

The largest excess of events as compared to isotropic expectations is observed from a region of the sky close to the location of the radio 
source Cen~A (galactic coordinates l=−50.5$^{\circ}$, b=19.4$^{\circ}$). The excess of events in circular windows around Cen~A with the
smallest isotropic chance probability corresponds to a radius of 18$^{\circ}$, which contains 12 events where 2.7 are expected on average 
if the flux were isotropic. By contrast, the region around the Virgo cluster is densely populated with galaxies but does not have an
excess of events above isotropic expectations. In particular, a circle of radius 20$^{\circ}$ centred at the location of M87 (galactic
coordinates l=76.2$^{\circ}$, b=74.5$^{\circ}$) does not contain any of 58 events. However, this is a region of relatively low exposure 
for the Pierre Auger Observatory and only 1.2 event is expected on average with the current statistics if the flux were isotropic.

\section{Plans for Future} 

First of all we can except more data coming from the current experiments. There are also some extensions of current 
observatories into lower energies, i.e. below $10^{18}$~eV, which are already in progress. The reason for these low energy extensions 
is making the comparison with the results from experiments (such as KASCADE-Grande \cite{Antoni,Bertaina}), which overlap with the direct 
measurements of cosmic rays. There are already several projects under construction or already in testing phase. At the Pierre 
Auger Observatory there are three high-elevation fluorescence telescopes (HEAT), which can be tilted and used to observe showers 
at energies down to $10^{17}$~eV \cite{Kleifges}. The other enhancement is AMIGA (Auger Muons and Infill for the Ground Array) 
which is formed by 85 pairs of a water Cherenkov tank and a buried muon counter arranged at 750~m and 433~m array spacings \cite{Platino}.
In other working experiment the Telescope Array Low Energy (TALE) extension is designed to observe cosmic rays down to energies 
of $3\times10^{16}$~eV \cite{Nonaka}.

The progress in next year can be expected also from the laboratory measurements of the fluorescence yield. The absolute measurement
of this quantity is expected particularly from the experiment AirFLY \cite{Bohacova}. Adopting this result will lead to 
a significant improvement of the energy reconstruction of all measured data.

There are a number of worldwide efforts to develop and establish new detection techniques that promise a cost-effective extension of currently 
available apertures to even larger dimensions. The detection of radio emission induced by ultra-high-energy cosmic rays hitting the Earth's
atmosphere is possible because of coherent radiation from the extensive air shower at radio frequencies. This radiation, which is emitted 
by secondary particles created in the air shower, can be measured with simple radio antennas. Radio detectors, like LOPES \cite{Haungs}
and CODALEMA \cite{Ardouin} produce promising results at energies beyond $10^{17}$ eV. 
The project AERA (Auger Engineering Radio Array) is planned at the Pierre Auger Observatory and it will have a dimension of about 20~km$^2$ \cite{Berg}.

The northern part of the Pierre Auger Observatory is planned to be built in the northern hemisphere. In 2005 the site was chosen
in southeast Colorado, USA. It will consist of an array of $4\ 000$ water tanks deployed on an area of $20\ 000$~km$^2$, arranged 
at $\sqrt 2$ miles array spacing on a square grid. The tanks will be observed by a set of fluorescence telescopes arranged in several 
locations. Research and development works on a small surface detector array, atmospheric monitoring, electronics and communication systems have 
already started and the construction is planned nowadays to start in the year 2011. Assuming the same UHECRs flux in the north sky there will be
detected approximately 180 events above the GZK energy per year.

Even more ambitious project is planned to take place at the International Space Station. Air shower tracks will be measured from this station 
in very large volume of the Earth's atmosphere. The key element of the sensor of JEM-EUSO (Japanese Experiment 
Module - Extreme Universe Space Observatory) will be a wide-field and fast large-lense telescope \cite{EUSO}. JEM-EUSO is planned to be 
launched before year of 2015.

\section{Conclusions} 

A significant progress in the study of UHECRs has been achieved by the data measured by the Pierre Auger Observatory.
The observatory is routinely measuring now and a release of new results can be expected soon. There will be also results from the Telescope 
Array during next years. Moreover, a comparison between the results measured in the southern and northern hemisphere by one experiment may 
be possible in close future. 

Up to now the suppression of the cosmic ray flux together with the anisotropy of arrival directions have been observed 
at the highest energies by the Pierre Auger Observatory. The flux of GZK photons and neutrinos should be reached in few years.
Moreover, the cross section between UHECRs and air can be calculated if the type of primary particle will be found.
All these results together with the observed arrival directions will hopefully lead to the finding of the sources of the most
energetic particles in the universe.


\end{document}